\documentclass[graybox,natbib,nosecnum]{svmult}
\bibpunct{(}{)}{;}{a}{}{,} 

\pdfoutput=1   

\usepackage{amsmath}
\usepackage{mathptmx}       
\usepackage{helvet}         
\usepackage{courier}        
\usepackage{type1cm}        

\usepackage{makeidx}         
\usepackage{graphicx}        
\usepackage{multicol}        
\usepackage[bottom]{footmisc}
\usepackage[normalem]{ulem}	
\usepackage{hyperref}  
\usepackage{epigraph}
\usepackage{soul}   

\makeindex             

\begin{document}

\title*{Exoplanet Catalogues}
\author{Jessie Christiansen}
\institute{IPAC, Mail Code 100-22, Caltech, 1200 E. California Blvd. Pasadena, CA 91125 \\ \email{jessie.christiansen@caltech.edu}}

\maketitle

\epigraph{All science is either physics or stamp collecting.}{\textit{Ernest Rutherford}}

\abstract{One of the most exciting developments in the field of exoplanets has been the progression from `stamp-collecting' to demography, from discovery to characterisation, from exoplanets to comparative exoplanetology. There is an exhilaration when a prediction is confirmed, a trend is observed, or a new population appears. This transition has been driven by the rise in the sheer number of known exoplanets, which has been rising exponentially for two decades \citep{Mamajek2016}. However, the careful collection, scrutiny and organisation of these exoplanets is necessary for drawing robust, scientific conclusions that are sensitive to the biases and caveats that have gone into their discovery. The purpose of this chapter is to discuss and demonstrate important considerations to keep in mind when examining or constructing a catalogue of exoplanets. First, we introduce the value of exoplanetary catalogues. There are a handful of large, online databases that aggregate the available exoplanet literature and render it digestible and navigable---an ever more complex task with the growing number and diversity of exoplanet discoveries. We compare and contrast three of the most up to date general catalogues, including the data and tools that are available. We then describe exoplanet catalogues that were constructed to address specific science questions or exoplanet discovery space. Although we do not attempt to list or summarise all the published lists of exoplanets in the literature in this chapter, we explore the case study of the NASA \emph{Kepler} mission planet catalogues in some detail. Finally, we lay out some of the best practices to adopt when constructing or utilising an exoplanet catalogue.}

\section{Why catalog exoplanets?}
\label{sec:value}

My children's STEM-focussed daycare centre recently spent a week concentrating on pets. This included the creation of a histogram to display on the wall of the number and types of pets in the class\footnote{Chickens were the surprising winner, with cats a respectable second.}. Children are natural scientists, a fact exploited and delighted in by our daycare, and this behaviour---cataloguing, grouping, comparing---is one of a scientist's most basic skills for understanding new phenomena. The children and I therefore take exception to Rutherford's characterisation of cataloguing as `lesser' science; in fact I would argue some of our most fundamental understanding of the universe around us could not have emerged without well-crafted catalogues, from Darwin's finches to Hubble's expansion of the universe. Catalogues function as a way to collect, organise, preserve and serve hard-won scientific data and derived results.

With a well-curated sample, one can make comparisons and draw conclusions. It should surprise no-one that after the discovery of 51 Peg b, the first exoplanet around a `normal' star, \citep{Mayor1995}, astronomers waited for only five additional planets to be discovered before performing the first comparative analysis \citep{Mazeh1997}. Comparing the new exoplanets to the solar system planets and the previously announced pulsar planets \citep{Wolszczan1992}, they observed that lower mass companions ($<$5~M$_{\rm Jup}$) appeared to reside on circular orbits, where the more massive companions had more eccentric orbits. Exoplanets could be grouped into two types---the era of comparative exoplanetology had begun. 

\section{General Catalogues}
\label{sec:general}

The large, general-use catalogues provide a comprehensive view of the state of the exoplanet field. They are very useful for providing a ready list of parameters for almost any known exoplanet (barring small differences in planet criteria, detailed in Table \ref{tab:general}). They also provide a list of references for each planet, pointing users back to the source of the information which can provide much-needed scientific context, and additional parameters that are not recorded in the catalogues. If multiple parameter sets are available for each planet, the catalogues can provide an historical archive of the knowledge of the planet parameters as they evolve (and typically improve) with time. These large aggregate catalogues are also useful for identifying and examining the broader population of exoplanets. However, particularly with this latter case, caution must be exercised. The papers from which the exoplanet parameters are drawn in these aggregate catalogues will have, for instance, different statistical thresholds, different reduction techniques, different modeling procedures, and different uncertainty philosophies. In order to perform robust population analyses, users are cautioned to examine carefully the selection effects and biases in the creation of the catalogue.

There are three large, online catalogues that are regularly updated. These are the Exoplanets Encyclopaedia\footnote{Exoplanets Encyclopaedia: \url{http://exoplanets.eu/}}, the NASA Exoplanet Archive\footnote{NASA Exoplanet Archive: \url{https://exoplanetarchive.ipac.caltech.edu/}}, and the Open Exoplanet Catalog\footnote{Open Exoplanet Catalog: \url{http://www.openexoplanetcatalogue.com/}}. Each has its own criteria for inclusion, which results in different total numbers of exoplanets listed; these are summarised in Table \ref{tab:general}. 

\begin{table}
\caption{The contents of the large, online exoplanet catalogues.}
\label{tab:general}
\begin{tabular}{llll}
\hline
Catalog & Mass criteria & Confidence criteria & Number of planets$^{\dagger}$\\
\hline
Exoplanet Encyclopaedia & $M_p-1\sigma<60M_{\rm Jup}$ & Submitted paper, conference talk & 3741\\
NASA Exoplanet Archive & $M_p<30M_{\rm Jup}$ & Accepted, refereed paper & 3704\\
Open Exoplanet Catalog & None listed & Open-source & 3504\\
\hline
\end{tabular}\\
$^{\dagger}$: as of February 27th, 2018. 
\end{table}

\subsubsection{Exoplanet Encyclopaedia}

\begin{figure}
\includegraphics[scale=.35]{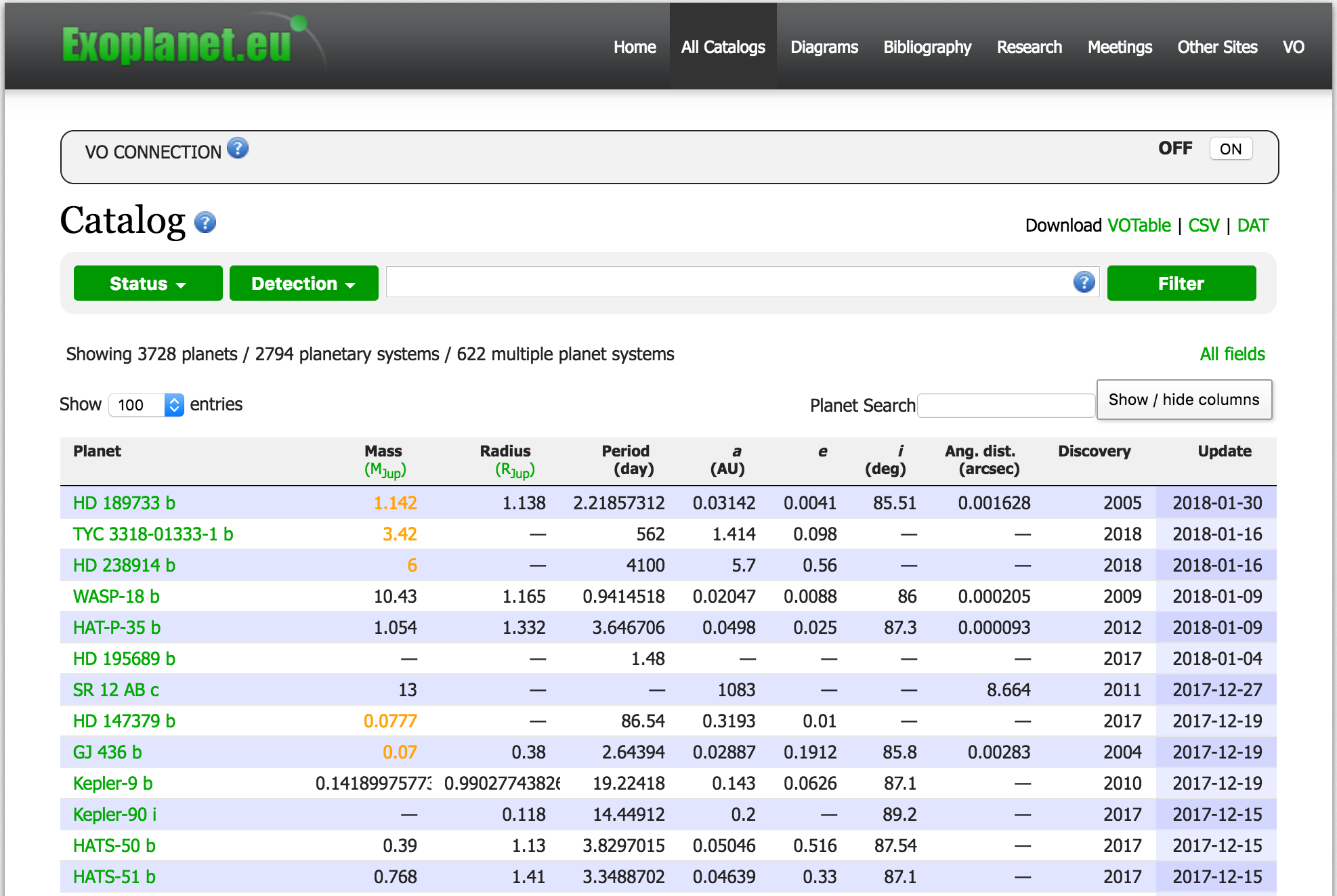}
\caption{A portion of the Exoplanet Encyclopaedia catalog, showing 3728 planets as of February 2nd, 2018. The full table can be filtered, sorted, and downloaded.}
\label{fig:ee}
\end{figure}

\begin{figure}
\begin{center}
\includegraphics[scale=.5]{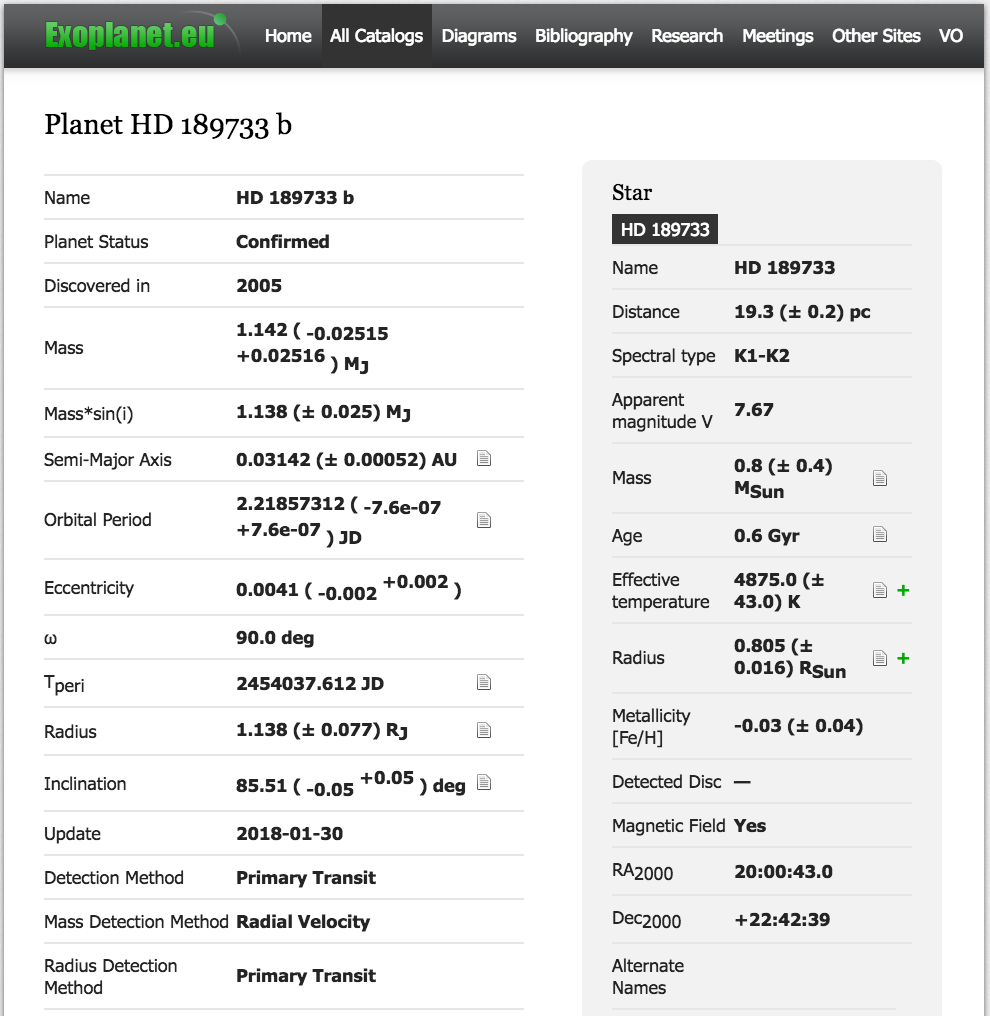}
\caption{A portion of the HD 189733 b overview page at the Exoplanet Encyclopaedia catalog.}
\label{fig:ee-overview}
\end{center}
\end{figure}

The Exoplanet Encyclopaedia \citep{Schneider2011} is maintained by Fran\c coise Roques and Jean Schneider at the Observatoire de Paris. It is updated rapidly (2--3 times per week) and includes all exoplanet announcements, for instance those announced at conferences or in submitted papers posted to the arXiv. Therefore, it typically has the highest number of catalogued exoplanets. The main catalog is presented as a large, interactive table, a preview of which is shown in Figure \ref{fig:ee}. A summary of the planet parameters is presented, and the table can be filtered, sorted, and downloaded in multiple formats. Each planet name in the catalog is a link to the overview page for that planet; Figure \ref{fig:ee-overview} shows an example overview page for HD 189733 b. The overview page lists one set of parameters for the planet, which can be drawn from multiple references. As a result, the parameters do not necessarily represent a physically consistent planet model, but can be a more complete set of parameters than is presented in any one reference. It also includes a very useful, comprehensive set of all published papers that contain references to a given planet, and a short set of remarks from the science curators summarising selected published results that are not obvious from the table of parameters (see Figure \ref{fig:ee-overview} for some example remarks). 

The Exoplanets Encyclopaedia also includes several useful tools that make use of the catalogue data. One is an interactive plotting tool\footnote{\url{http://exoplanet.eu/diagrams/}} with the data pre-loaded; users can customise the parameters that are plotted and the appearance of the plot, and download the resulting figure. Another tool, the Observability Predictor, is linked from the planet overview pages. For a given set of planet parameters and a defined observation window will plot the observable orbital parameters of the system, such as angular separation or star-planet distance. For many multi-planet systems, there is a link to a calculation of the Angular Momentum Deficit (AMD) stability value, which can be used to classify the stability of the planetary system; they plan to extend this to all multi-planet systems soon (F. Roques, private communication). Finally, the Exoplanets Encyclopaedia also maintains a highly used list of Future Meetings on Extrasolar Planets\footnote{\url{http://exoplanet.eu/meetings/}}, which provides a comprehensive list of the conferences and workshops around the world that pertain to exoplanet studies and results.

\subsubsection{NASA Exoplanet Archive}

\begin{figure}
\includegraphics[scale=.26]{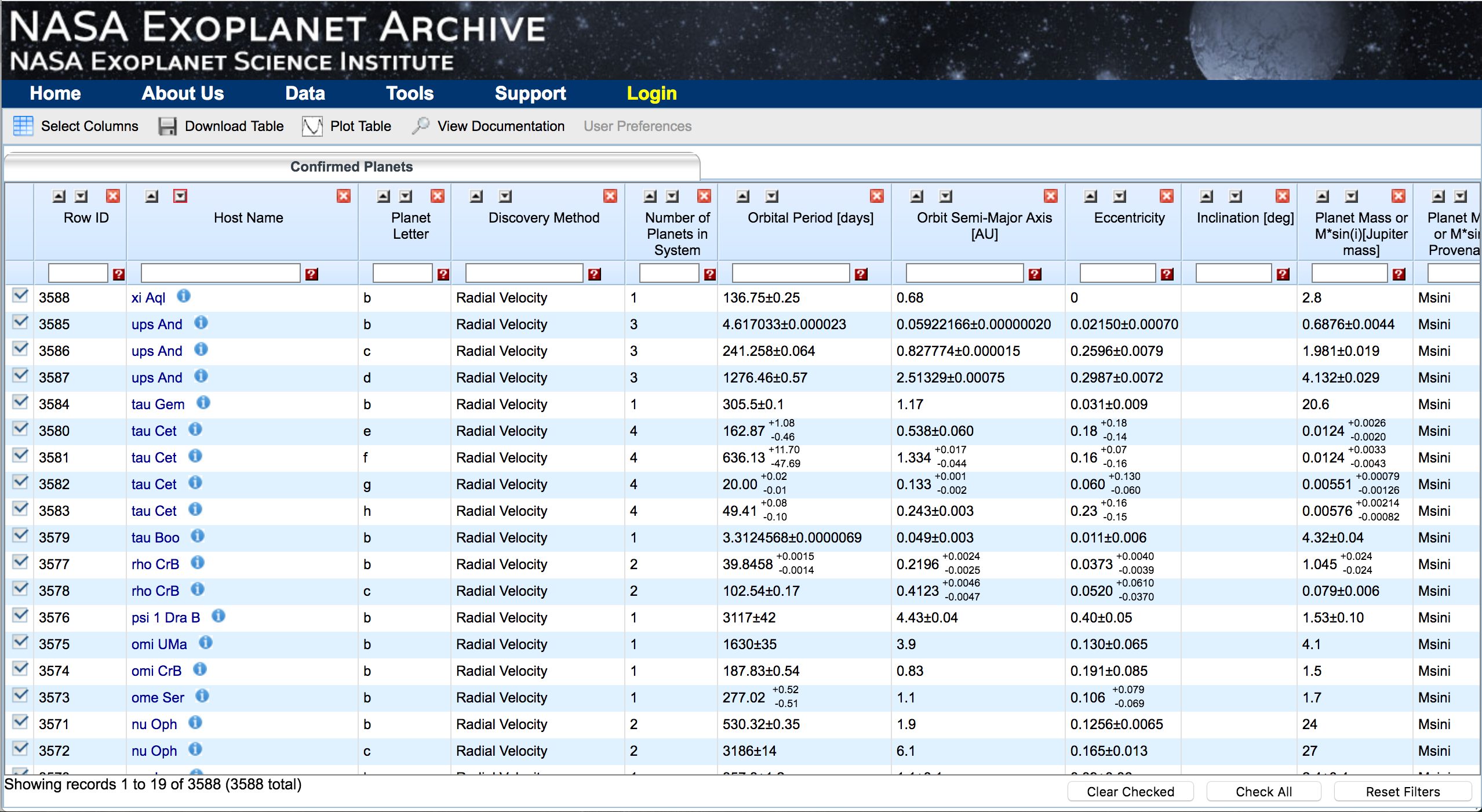}
\caption{A portion of the NASA Exoplanet Archive confirmed planets table, showing 3588 planets as of February 2nd, 2018. The full table can be filtered, sorted, and downloaded.}
\label{fig:nea}
\end{figure}

The NASA Exoplanet Archive is maintained by the NASA Exoplanet Science Institute \citep{Akeson2013}. It is updated weekly, and includes only those exoplanets announced in refereed publications\footnote{The recent resurrection of the Research Notes of the American Astronomical Society, which are moderated by an editorial board but not refereed, may lead to a re-evaluation of this position depending on its eventual use by the community.} The confirmed planets table is shown in Figure \ref{fig:nea}; like the Exoplanets Encyclopaedia the table is interactive---parameter columns can be added or removed, filtered and sorted, and the table can be downloaded in multiple formats. 

Each row in the Exoplanet Archive confirmed planets table shows the `default' set of parameters for that exoplanet. This is a concept that deserves some discussion, since it represents one of the philosophical decisions encountered on construction of such a table. As described above, the Exoplanets Encyclopaedia presents parameters from multiple references, which provides a complete picture of those parameters which have been derived, but may not be a physically consistent set of parameters. In contrast, the Exoplanet Archive chooses a physically consistent set of parameters from a single analysis, which may be incomplete depending on the specific parameters published by the authors. This `default' set is chosen to balance the completeness and precision of the published parameters, as well as the potential staleness of the ephemeris. However, this can lead to seemingly confusing gaps in the confirmed planets table, where important parameters which have been measured for high-profile exoplanets (e.g. planet radius) appear blank. In order to address this, the Exoplanet Archive maintains multiple published parameters sets per planet, described in more detail below.

\begin{figure}
\includegraphics[scale=.26]{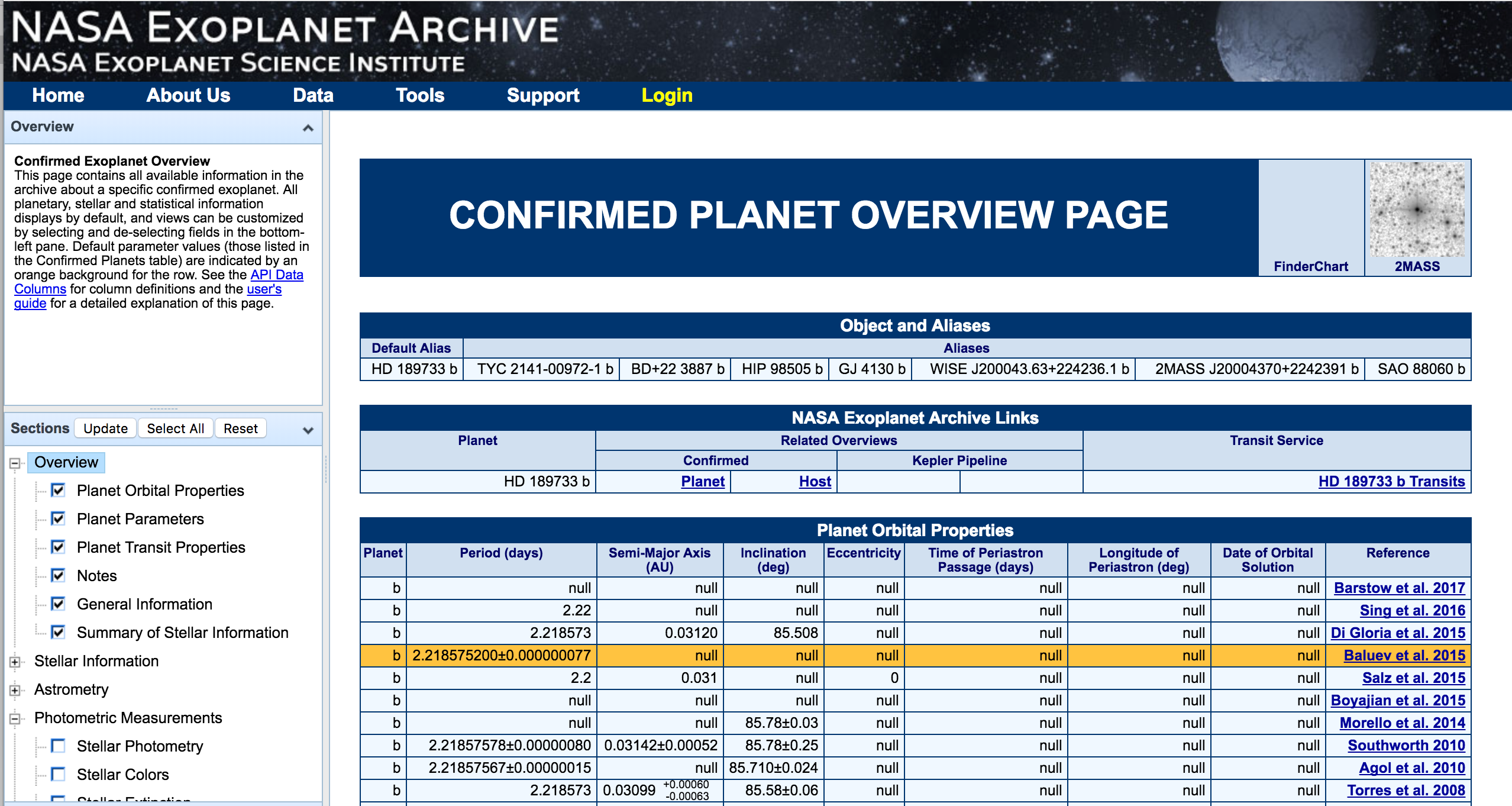}
\caption{A portion of the HD 189733 overview page at the NASA Exoplanet Archive. Where multiple parameter sets are available, the row highlighted in yellow denotes the `default' set of parameters shown in the confirmed planets table (see text for details). Note also that the left-most panel contains a full list of tables that can be interactively made visible or invisible; much of the extended data is hidden by default.}
\label{fig:nea-overview}
\end{figure}

In the Exoplanet Archive confirmed planets table, the blue icon next to each exoplanet name (seen in Figure \ref{fig:nea}) contains links to a variety of associated pages, including planet and host overview pages, the transit/ephemeris prediction service, and for NASA \emph{Kepler} mission planets, links to photometric time series and additional Kepler-specific overview pages. Figure \ref{fig:nea-overview} shows an example overview page for HD 189733. The Exoplanet Archive overview pages list multiple published parameter sets for each planets. Newly published parameter sets are compared to the extant values and ingested if they represent an improvement in the state of the knowledge of the exoplanet. Therefore the parameter displayed on the overview pages do not necessarily constitute a comprehensive set of the published parameters. The set of available exoplanet parameters can also be accessed via the `Search Extended Planet Data' link on the Exoplanet Archive homepage, under `Tools \& Services'. 

The Exoplanet Archive also houses over 40,000,000 light curves, largely from the UKIRT, SuperWASP and \emph{Kepler} missions, and including contributions from the CNES \emph{CoRoT} mission and several ground-based surveys. It is the archive for the \emph{Kepler} planet search products, including the planet candidate tables, data validation reports and summaries, and survey completeness and reliability products, and also maintains a list of the subsequent \emph{K2} mission planet candidates. It hosts tables of transmission and emission spectroscopy values that are compiled from the published literature.

Like the Exoplanets Encyclopaedia, the Exoplanet Archive deploys several tools that utilise the underlying data. One is a standard set of pre-generated, presentation-ready plots, which are automatically generated with each update. There is a transit/ephemeris prediction service for observation and proposal planning. In the service, users can generate lists of viewable events (transits, eclipses, quadrature or arbitrary phases), from a given Earth- or space-based location (or independent of observator location), in a given observation window, for either one of the existing exoplanet lists, a user-defined list, or an individual planet (including a blank template for custom ephemerides). There is a periodogram fitting tool, which for an uploaded time series can perform a Lomb-Scargle periodogram, a Box Least Squares fit, or a Plavchan periodogram, with customisable period ranges and steps. There is interactive version of the publicly available EXOFAST transit and radial velocity fitting tool \citep{Eastman2013}, which is supported by enough computing power to allow for Monte-Carlo Markov chain error analysis.There is the `Predicted Observables for Exoplanets' service, which includes predictions of both stellar habitable zones and also planetary signatures such as radial velocity semi-amplitude, the astrometric semi-amplitude, the transit depth, maximum separation on the sky. Finally, most of the interactive tables at the Exoplanet Archive are linked directly to an interactive plotting tool for viewing the data.

\subsubsection{Open Exoplanet Catalogue}

\begin{figure}
\includegraphics[scale=.26]{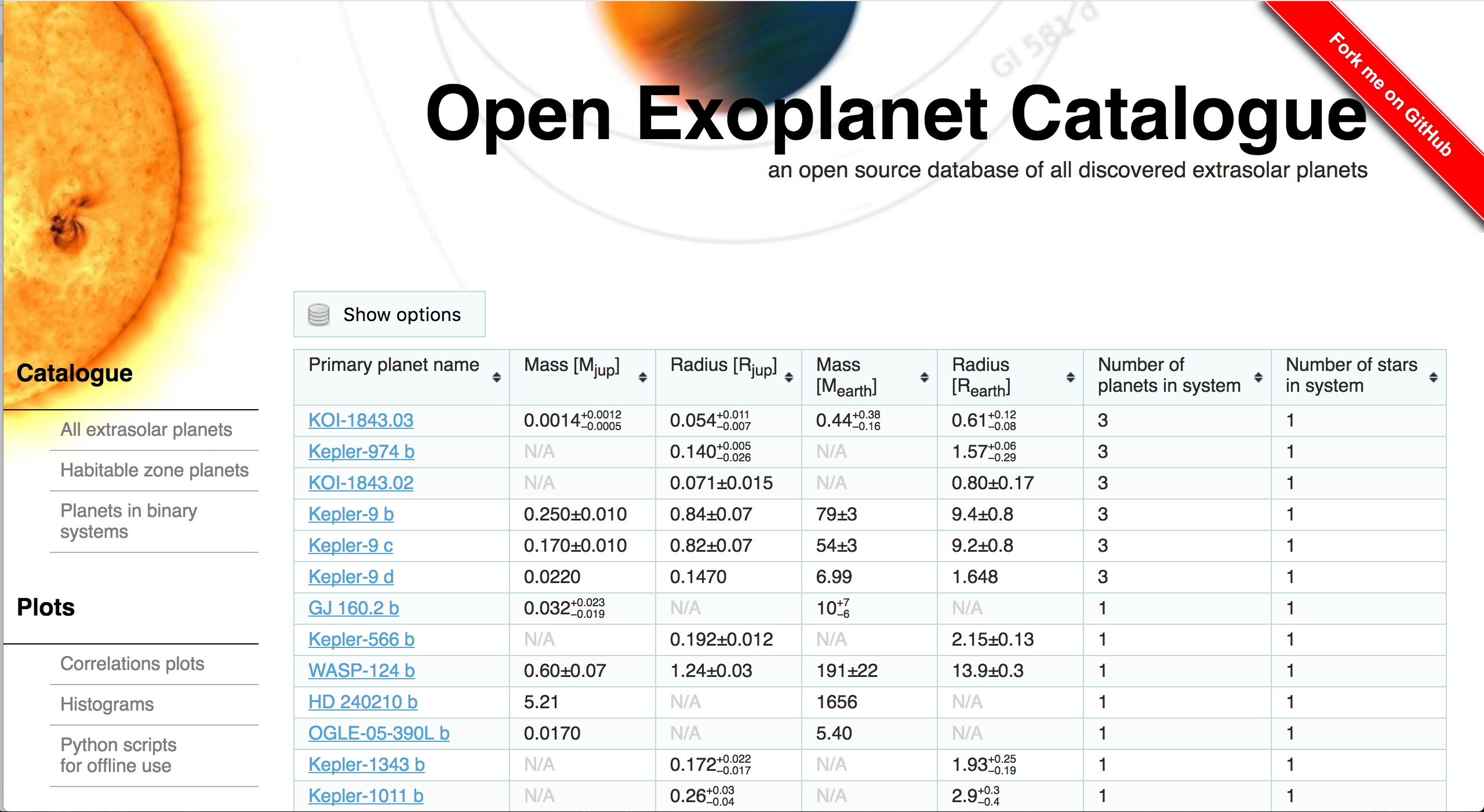}
\caption{A portion of the Open Exoplanet Catalog, which had 3504 planets as of February 2nd, 2018. The full table can be sorted and downloaded, and there are also limited filtering options.}
\label{fig:oec}
\end{figure}

The Open Exoplanet Catalogue is an open-source, decentralised database of exoplanets. Professional astronomers are encouraged to contribute updates and corrections, and as such the update schedule is sporadic but can be multiple times per week. It includes all exoplanets submitted to the database. There is a web interface to the full database, shown in Figure \ref{fig:oec}, but the primary use case is download through github\footnote{Open Exoplanet Catalogue: \url{https://github.com/OpenExoplanetCatalogue/open\_exoplanet\_catalogue/}}. This nimble approach allows updates and corrections (e.g. for typographical errors) to be made rapidly by any contributor via a pull request. The github commit messages for new or updated data are requested to contain the citation to the paper from which the data are extracted, for traceability.

\begin{figure}
\includegraphics[scale=.26]{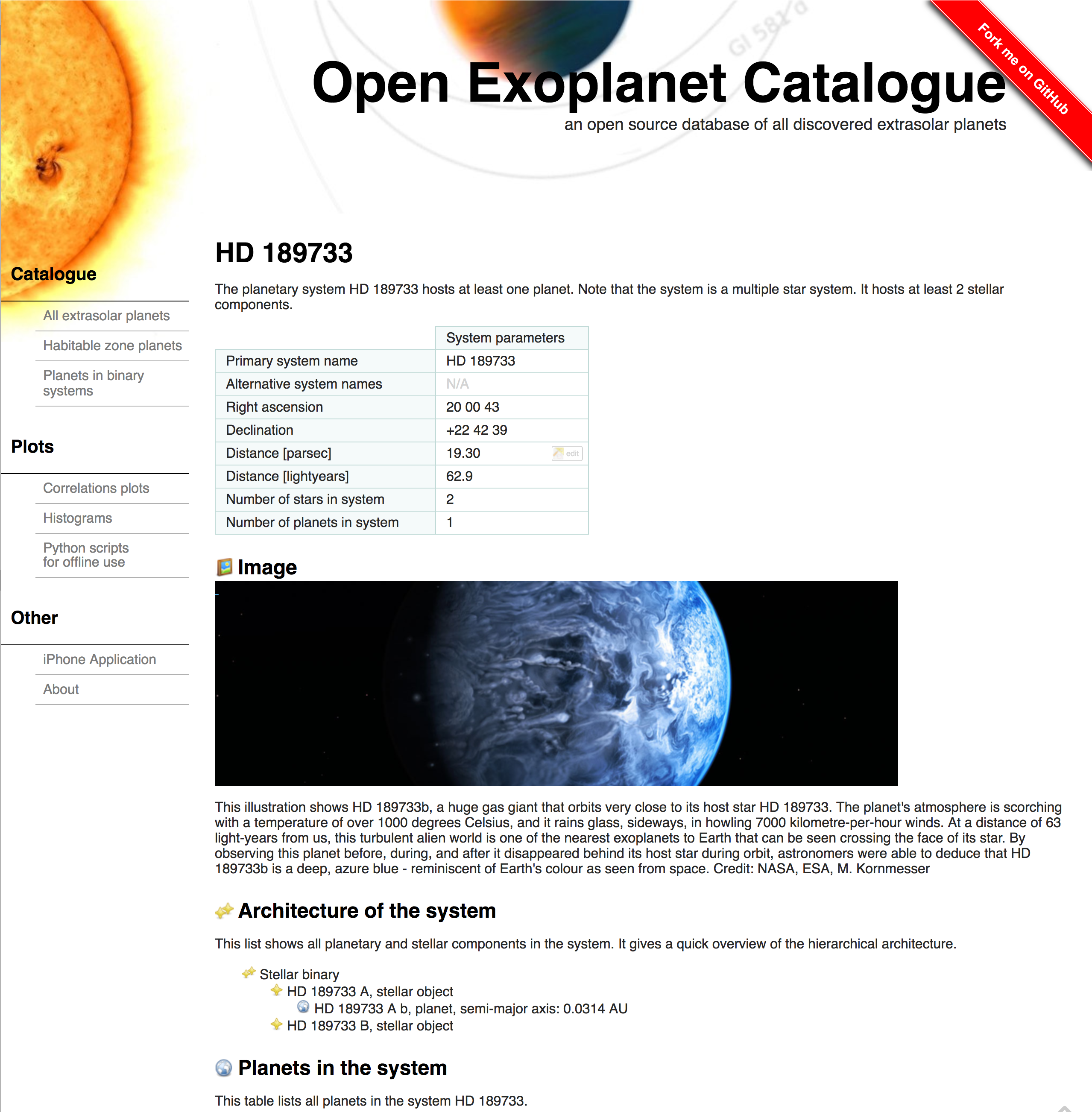}
\caption{A portion of the HD 189733 overview page at the Open Exoplanet Catalog.}
\label{fig:oec-overview}
\end{figure}

As for the previously described tables, the exoplanet names in the Open Exoplanet Catalogue interactive website table are links through to overview pages for each planet. Figure \ref{fig:oec-overview} shows the overview page for HD 189733. Each page contains a single set of parameters for each planet and star in the system, as curated by the contributors to the repository; the issue of parameter physical consistency versus completeness would need to be traced by examining the origin of each uploaded parameter on a case-by-case basis. The Open Exoplanet Catalogue overview pages contain some very useful automatically generated graphics which show the scale of the given exoplanet to the solar system planets, and also the scale of the orbit. They also handle all system architectures (multiple stars, multiple planets) elegantly with a section of the overview page dedicated to describing the architecture. 

The Open Exoplanet Catalogue also contains both in-browser tools, including an interactive plotter, and an iPhone app for mobile use. Also provided are a set of python scripts for offline interaction with the database.

It should be clear to the reader that each of these large, general archives has its own strengths, and its own use cases. For a final note, one interesting difference between the three catalogues described here is their treatment of free-floating planets: the Exoplanets Encyclopaedia and the Open Exoplanet Catalogue include them, whereas the Exoplanet Archive does not.

\section{Catalogues for specific science cases}
\label{sec:specific}

In addition to the large, general catalogues, there are various use cases that require more carefully curated lists of exoplanets, or that handle more niche exoplanet properties. We describe several of the most utilised, specialised catalogues here, covering a variety of science cases.

\subsubsection{Highly reliable planets}

Some of the earliest exoplanet catalogues were, by necessity, lists of radial velocity planets. One of the earliest published list of exoplanets that we could find is Table~3 of \citet{Butler2002}, which lists the 57 radial velocity planets with robust detections that had been published to the date of that paper. There are likely earlier tables that escaped our attention, but the Butler table has the distinction of going on to form the basis of the Exoplanet Orbit Database\footnote{Exoplanet Orbit Database: \url{http://www.exoplanets.org}} \citep{exoplanetsdotorg, exoplanetsdotorgtwo}. While this database originally focussed on robust radial velocity detections, it eventually expanded to include all known exoplanets with a robust orbit. It continued to maintain a high standard for inclusion, and the science curators performed independent checks of published parameters against available radial velocity curves and corrected them where necessary to ensure high accuracy. The database contains a single, highly vetted set of parameters for each planet. Unfortunately the database has not been actively updated since 2014, but there are plans to restart regular updates in the near future (J. Wright, private communication).

\subsubsection{Transiting planets}

One issue with aggregate catalogues discussed earlier is the lack of uniformity in analysis, which can make comparing planets problematic. The Transiting Exoplanet Parameter Catalog\footnote{TEPCAT: \url{http://www.astro.keele.ac.uk/jkt/tepcat/}} (TEPCAT), maintained by John Southworth at Keele University, addresses this in part by providing (amongst other features) a large, uniform re-analysis of many transiting exoplanet systems, as published in \citet{Southworth2010, Southworth2011, Southworth2012}. These and other robustly measured parameters of transiting planets are available for download in html, ASCII and CSV formats. The website also provides some nice pre-generated plots from the underlying transit planet data.

The Exoplanet Transit Database\footnote{Exoplanet Transit Database: \url{http://var2.astro.cz/ETD}} is a large repository of transit observations of transiting planets \citep{Poddany2010}, both extracted from the published literature and contributed by the community. It is maintained by the Variable Star Section of the Czech Astronomical Society. The database contained 6741 individual transits on 326 objects as of February 4th, 2018, each rated with a Data Quality index. The website also includes several tools, including interactive plotting of the transit data, a transit model-fitting tool, and a transit event prediction service.

\subsubsection{Habitable zone planets}

The Habitable Zone Gallery\footnote{Habitable Zone Gallery: \url{http://www.hzgallery.org/}} is maintained by Stephen Kane (UC Riverside) and Dawn Gelino (NASA Exoplanet Science Institute). For planets with a complete orbital solution, the catalogue calculates both the conservative and optimistic habitable zones for the host star, based on \citet{Kane2012}, and the percentage of the planet's orbit that it spends within those zones. The calculations are presented in tabular form, available for download in CSV, and in convenient graphical form for presentations, including both still images and animated movies for each planet's orbit. As of February 2nd, 2018, there are 92 planets listed with orbits entirely within the habitable zone.

\subsubsection{Exoplanet occurrence rates}

The primary science goal of the NASA \emph{Kepler} mission was to measure the occurrence rate of planets like the Earth around stars like the Sun. This measurement requires many ingredients, including a well characterised sample of stars, and a well understood catalogue of planet candidates detected around those stars. The mission gathered observations for four years from 2009 to 2013, and began producing lists of planet candidates almost immediately; the first catalogue was produced from only the first month of data \citep{Borucki2011a}. In total the mission published eight separate planet candidate catalogues, listed in Table \ref{tab:kepler}; the final six catalogues are available as interactive tables at the NASA Exoplanet Archive\footnote{Kepler Objects of Interest: \url{https://exoplanetarchive.ipac.caltech.edu/cgi-bin/TblView/nph-tblView?app=ExoTbls&config=koi}}. Each subsequent catalogue was produced with a longer observing baseline (excepting the final catalogue which was a re-analysis of the full four years), and each represented an increased understanding in the underlying properties of the listed planet candidates. 

Table \ref{tab:kepler} summarises the evolution in the catalogue properties. One important aspect is the degree of uniformity in the origin of the contents of each catalogue. Early catalogues contained aggregate lists of planet candidates compiled from multiple sources, including searches by eye of the data. For use in statistical occurrence rate calculations, later catalogues comprised only those planet candidates derived from a single source, typically a uniform re-processing of the data by the latest version of \emph{Kepler} pipeline. As a result, there are planet candidates in the earlier catalogues that, although they subsequently became confirmed or validated planets, are not present in later catalogues if they were not re-discovered by the corresponding version of the pipeline. Finally, as the analysis matured and the final data were obtained, attention turned to measuring the underlying completeness (false negative rate) and reliability (false positive rate) of the planet candidates in the catalogues. The final \emph{Kepler} mission planet candidate catalogue (Thompson et al. submitted) is the first for which the completeness and reliability are both quantitatively measured. As mentioned previously, the final completeness and reliability products are hosted at the NASA Exoplanet Archive\footnote{Kepler Completeness and Reliability Products: \url{https://exoplanetarchive.ipac.caltech.edu/docs/Kepler_completeness_reliability.html}}; see the documentation there for more details of the types of products available.

\begin{table}
\caption{Kepler Object of Interest$^{\dagger}$ catalogues.}
\label{tab:kepler}       
\begin{tabular}{lllllll}
\hline
Reference & Quarters$^{\ddagger}$ & Observing    & Total  & Uniform   & Completeness & Reliability \\
                  & included                        &  baseline     &   PCs  & selection & measured        & measured \\
\hline
\citet{Borucki2011a} & Q1M1 & 33.5~d  & 305 & No & No & No \\
\citet{Borucki2011b} & Q0--Q5$^{\clubsuit}$ & 13 months  & 1202 & No & No & No \\
\citet{Batalha2013} & Q1--Q6 & 16 months  & 2338 & No & No & No \\
\citet{Burke2014} & Q1--Q8 & 2 years  & 2738 & No & No & No\\
\citet{Rowe2015} & Q1--Q12 & 3 years  & 3697 & No & No & No\\
\citet{Mullally2015} & Q1--Q16 & 4 years  & 4175 & Yes & Yes & No\\
\citet{Coughlin2016} & Q1--Q17 & 4 years  & 4696 & Yes & Yes & Partial\\
Thompson et al. (submitted) & Q1--Q17 & 4 years  & 4034 & Yes & Yes & Yes\\
\hline
\end{tabular}\\
\footnotesize
$^{\dagger}$: Note that the definition of a `Kepler Object of Interest' evolved over the course of the mission; for simplicity here we list the contents of each catalog as they were defined at the time. \\
$^{\ddagger}$: A `quarter' of data is three continuous months of observations obtained at the same satellite orientation; \emph{Kepler} operated for seventeen quarters, with a short 10-day commissioning period referred to as Quarter 0. Q1M1 refers to month 1 of the quarter 1 observations.\\
$^{\clubsuit}$ KOIs were identified using the Q0--Q2 data, but were characterised using the Q0--Q5 data.
\end{table}

%

\section{Best practices}
\label{sec:best}

What makes a good catalogue? It does depend on your use case, but here are some general properties to maximise utility and longevity.

\begin{itemize}
\item Catalogues should list objective, documented criteria for inclusion. For instance, what is an `exoplanet'? There is no formal IAU designation. Is there a mass limit? Is a host star required? Is the catalogue under consideration permissive (e.g. the Exoplanets Encyclopaedia) or restrictive (e.g. the Exoplanet Orbit Explorer database)? Understanding the reasons for inclusion is important for evaluating whether a given exoplanet catalogue meets the science needs of the user. As a corollary, documenting deliberate exclusions from a catalogue (and the reasons why) is useful if that user is unable to locate their selected exoplanet.
\item Catalogues should make citations for externally sourced data readily available. Almost all the catalogues described in this chapter comprise at least some parameters from published sources; traceability of these sources is essential for providing credit, context, and accuracy.
\item Catalogue contents should be described with clear and complete meta-data. This includes full parameter lists, definitions (including units) and descriptions. An important but somewhat subtle point is that, for machine-readability, the definition of `null' values should be consistent and clearly defined.
\item Catalogue data should use standard units and standard conversions where they have been established. The IAU provides a list of standard conversion constants, for example for converting from Jupiter radii to solar radii, which should be used where possible\footnote{Resolution B3 at \url{https://www.iau.org/static/resolutions/IAU2015_English.pdf}}.
\item The catalogue data should be available for download in standard formats (CSV, ASCII, VOTable, IPAC table, etc.)
\item Finally, the catalogue data should be backed up with redundant, off-site copies. Sometimes the main power feed to the building shorts in its underground conduit at 2am and takes down the whole city block for 15 hours, which is longer than your UPSes can handle, and you just can't predict these things. Back-up your data.
\end{itemize}

\section{Conclusion}

It is a tremendous time to be working in exoplanets, as will be well demonstrated throughout this textbook. The rapid pace of discovery, matched with a shifting philosophy into open-source data and software packages, mean that convenient, accessible online databases and tools are flourishing. This chapter has summarised several of the large, regularly updated aggregator websites, and explored some of the catalogues compiled to address specific science cases, including the specific case study of the \emph{Kepler} mission planet candidate catalogues. Finally, some best practices for constructing a useful, archival-quality exoplanet catalogue are outlined.

It is exhilarating to look to the future from our current vantage point. As of writing, there are 3500--3700 exoplanets (depending on which catalogue you use). The NASA \emph{TESS} mission will launch in several months, and is predicted to find another 23,000 planets \citep{Sullivan2015}. The NASA \emph{WFIRST} mission is slated for launch in the mid-2020's, and is predicted to yield $\sim$100,000 more \citep{Spergel2015,Montet2017}. The ESA \emph{Gaia} and {PLATO} missions will increase the totals still further in the next decade. Exoplanet catalogues, as systems for organising and making easily accessible these exoplanets and their attendant data, will find new ways to deal with the increased loads, and develop new tools to exploit the increasingly comprehensive data sets. As the pace of data ingestion and subsequent maintenance outstrips what an individual researcher can sustain, the large online catalogues will become ever more important and utilised. Tantalisingly, the new discoveries about the populations of exoplanets that will arise from the study of carefully curated exoplanet catalogues can, as yet, only be imagined.

\section{Acknowledgements}


\end{document}